\documentclass[12pt]{iopart}
\usepackage{url}
\usepackage{hyperref}

\usepackage{graphicx}

\newcommand{\bef}{\begin{figure}}
\newcommand{\eef}{\end{figure}}

\newcommand{\be}{\begin{equation}}
\newcommand{\ee}{\end{equation}}
\newcommand{\bea}{\begin{eqnarray}}
\newcommand{\eea}{\end{eqnarray}}

\begin{document}


\title{Unfolding of event-by-event net-charge distributions in heavy-ion collision}
\author{P.~Garg$^1$, D.~K.~Mishra$^2$, P.~K.~Netrakanti$^2$, A.~K.~Mohanty$^2$ and B.~Mohanty$^3$}
\address{$^1$ Department of Physics, Banaras Hindu University, Varanasi 221005, India\\
         $^2$ Nuclear Physics Division, Bhabha Atomic Research Center, Mumbai 400094, India \\
         $^3$ School of Physical Sciences, National Institute of Science Education and Research, Bhubaneswar 751005, India}

\begin{abstract}

We discuss a method to obtain the true event-by-event net-charge multiplicity 
distributions from a corresponding measured distribution which is subjected to
detector effects such as finite particle counting efficiency. The
approach is based on the Bayes method for unfolding of distributions. 
We are able to faithfully unfold back the measured distributions to match with 
their corresponding true distributions obtained for a widely varying underlying 
particle production mechanism, beam energy and collision centrality. 
Particularly the mean, variance, skewness, kurtosis, 
their products and ratios of net-charge distributions from the event generators 
are shown to be successfully unfolded from the measured distributions constructed 
to mimic a real experimental distribution. We demonstrate the necessity to account
for detector effects before associating the higher moments of net-charge distributions
with physical quantities or phenomena. The advantage of this approach being that one 
need not construct new observable to cancel out detector effects which loose their
ability to be connected to physical quantities calculable in standard theories.

\end{abstract}
\maketitle
\section{Introduction}
\label{intro}
Higher moments of the event-by-event distribution of conserved quantities like net-charge,
net-baryon number and net-strangeness in heavy-ion collisions have been found to be useful 
observable to characterize the system formed in the collisions
~\cite{Aggarwal:2010wy}. Higher moments have been shown to be related to the correlation
length~\cite{Stephanov:2008qz} and susceptibilities ~\cite{Bazavov:2012jq,Cheng:2008zh} of
the system and hence can be used to look for signals of phase transition and 
critical point~\cite{Stephanov:2011pb,Asakawa:2009aj,Hatta:2003wn}.
They have also been shown to be useful for studying the bulk QCD thermodynamics at high
temperature~\cite{Gupta:2011wh}. Specifically, proposals have been made to extract the
freeze-out properties of the system using higher moments of net-charge and net-baryon
number distributions, in a way very similar to that done using the particle yields and
ratios~\cite{Gavai:2010zn,Karsch:2010ck,Friman:2011pf}.

Any experimental measurement is susceptible to the effects such as the finite acceptance,
finite efficiency of counting the number of particles produced in the collisions and other
background effects~\cite{Abelev:2008ab}. It is almost impossible to know some of these
quantities for each event so as to correct for the effects in an event-by-event
distribution. Hence most of the experimentally measured event-by-event distributions are 
presented without these corrections~\cite{Aggarwal:2010wy,Appelshauser:1999ft,Aggarwal:2001aa,Adcox:2002mm}.
These corrections are carried out on an average level for reporting the yields 
of the produced particles (typically the first moment of the multiplicity distributions)
~\cite{Abelev:2008ab}. Comparison of uncorrected experimental event-by-event
distributions to theoretical calculations needs to be done carefully. For example, using
the corrected mean multiplicities to explain the uncorrected measured event-by-event
distributions could lead to different conclusions~\cite{BraunMunzinger:2011dn,bmcpod2011}.

Judicious construction of event-by-event observables have been proposed to cancel out
detector effects to first order~\cite{Mrowczynski:1999un,Voloshin:1999yf,Bialas:2007ed,Pruneau:2002yf}.
However, while making these constructs, one may sometimes loose the ability to compare them 
to the theoretically calculated quantities in order to extract meaningful physical insights.
That introduce additional complexities which makes it difficult for a proper physical
interpretation of the observable.  As an example, the moments of the multiplicity distribution 
of conserved quantities can be shown to be proportional to correlation length ($\xi$) of the system. 
The variances ($\sigma^2$ $\equiv$ $\left\langle (\Delta N)^2 \right\rangle$; $\Delta N = N - M$; $M$ is
the mean) of these distributions are related to $\xi$ as $\sigma^2$ $\sim$ $\xi^2$,
skewness (${\it {S}}$ = $\left\langle (\Delta N)^3 \right\rangle/\sigma^{3}$) goes as
$\xi^{4.5}$ and kurtosis ($\kappa$ = [$\left\langle (\Delta N)^4
\right\rangle/\sigma^{4}$] - 3) goes as $\xi^7$~\cite{Stephanov:2008qz}. Their product
such as $\kappa$$\sigma^2$ are related to the ratio of fourth order ($\chi^{(4)}$) to
second order ($\chi^{(2)}$) susceptibilities~\cite{Bazavov:2012jq,Cheng:2008zh}.
Where $\chi^{(2)} = \frac{\left\langle (\Delta N)^{2}\right\rangle}{VT}$, {\it V} is
the volume, and $\Delta N$ could be the net-baryon number or net-charge number.
In order to cancel out the acceptance and efficiency effects to first order for these
observables, constructs such as normalized factorial moments (defined later) can be made. 
The factorial moments of a particular order however become complicated function of 
lower order moments. Thereby making their interpretation difficult in terms of physical 
observables such as $\xi$ or $\chi$ calculated in a standard theory.

Here we give a simple calculation to illustrate our point of view. Let $N$ represents the
produced multiplicity and $n$ being the actually measured multiplicity in an experiment.
We parametrize the detector response in the experiment by a binomial probability
distribution function given by,

\begin{equation}
B(n:N,\epsilon) = \frac{N!}{n!(N-n)!} \epsilon^{n} (1-\epsilon)^{N-n},
\end{equation}
where $\epsilon$ is the particle counting efficiency.

We further consider that the produced multiplicity follows probability distribution
function $P(N)$, and that for measured distribution is $P(n)$. Then the mean of measured
multiplicity distribution $\langle n \rangle$ can be related to the mean of the actually 
produced multiplicity distribution as,

\begin{eqnarray}
\langle n \rangle = \int n P(n)dn = \int n dn \int B(n\mid N) P(N) dN \\ \nonumber
= \int P(N)dN \int B(n\mid N) n dn = \epsilon \int P(N) N dN = \epsilon \langle N \rangle.
\end{eqnarray}
Similarly it can be shown that,
\begin{equation}
\langle n^2 \rangle = \epsilon (1 - \epsilon) \langle N \rangle + \epsilon^{2} \langle N^2 \rangle.
\end{equation}

Now let us suppose that we can correct event-by-event particle counting efficiency,
the variance of the resultant measured distribution can be shown to be,  
\begin{equation}
\sigma^{2}(n/\epsilon) = \frac{1 - \epsilon}{\epsilon} \langle N \rangle +  \sigma^{2}(N)
\end{equation}
We find that the variance of $n/\epsilon$ is not equal to the variance of $N$ even though the
mean of $n/\epsilon$ is equal to the mean of $N$. Similar derivations and conclusions can
be done for higher order moments.

Alternatively, one can construct second order factorial moments such as
\begin{equation}
\frac{\langle n(n-1) \rangle}{\langle n \rangle^{2} } = \frac{\epsilon^{2} \langle N(N-1) \rangle}{\epsilon^{2}\langle N \rangle^{2}} = \frac{\langle N(N-1) \rangle}{\langle N \rangle^{2} },
\end{equation}
and  the fourth order factorial moment as, 
\begin{equation}
\frac{\langle n(n-1)(n-2)(n-3) \rangle}{\langle n \rangle^{4} } =
\frac{\langle N(N-1)(N-2)(N-3) \rangle}{\langle N \rangle^{4} }.
\end{equation}
These are found to be independent of efficiency effects. In these we 
also assume that $\epsilon$ does not vary event-by-event.

However a closer look at these construct will reveal that,
\begin{equation}
\frac{\langle n(n-1) \rangle}{\langle n \rangle^{2} } =
\frac{\sigma^{2}}{\langle n \rangle^{2}} - \frac{1}{\langle n \rangle} + 1
\end{equation}
and
\begin{eqnarray}
\frac{\langle n(n-1)(n-2)(n-3) \rangle}{\langle n \rangle^{4} } =
11 \frac{\sigma^{2}}{\langle n \rangle^{4}} + \frac{1}{\langle n \rangle^{2}} \\
\nonumber 
- 6{\it S}\frac{\sigma^{3}}{\langle n \rangle^{4}} -
18 \frac{\sigma^{2}}{\langle n \rangle^{3}} + \kappa \frac{\sigma^4}{\langle n
\rangle^{4}} \\ \nonumber 
+ 4 {\it S} \frac{\sigma^{3}}{\langle n \rangle^{3}} - 3 \frac{\sigma^{2}}{\langle n
\rangle^{2}} - 2 - \frac{6}{\langle n \rangle^{3}}.
\end{eqnarray}
While trying to remove the detector effects we have arrived at constructs which loose the
purity of moments or become involved functions of lower order moments. Thereby making it
difficult to directly connect to physical observables such as susceptibilities or their
ratios which can give important insight to the bulk properties of the system formed in
heavy-ion collisions.

Keeping in view the importance of higher moments of multiplicity distributions to
characterize the system formed in heavy-ion collisions, it is necessary to have a
proper way to compare the measurements and theory calculations. At the same time 
ensure that experimental artifacts like acceptance and particle counting efficiency
are removed. In this paper, we propose an approach based on unfolding of the measured 
(actually measured in experiments) multiplicity distribution to get back the true (actually produced in the collisions)
distributions produced in the collisions. Such a method  seems to work only if the detector
response can be satisfactorily modeled and the statistics is large enough. 

The paper is organized as follows. In the next section we discuss the event generators used 
in this study. In section 3 we discuss method of unfolding.
In section 4 we present the results for the moments of the net-charge distribution as a function of collision
centrality (defined in terms of the number of participating nucleons, $N_{\rm {part}}$). 
A brief discussion of the limitations of the approach is also presented. 
Finally we summarize our study in section 5.

\section{Event generators}

In this study we have used two event generators HIJING~\cite{Gyulassy:1994ew} (version 1.37) 
and THERMINATOR~\cite{Kisiel:2005hn} (version 2.0). 
They provide the possibility of different probability distribution for charged particle
multiplicity, to study our proposal.
While HIJING distributions are based on the physics due to QCD inspired models for multiple jet production,
the THERMINATOR distributions are based on systems in thermodynamical equilibrium. 
The further details of the models can be found in Ref.~\cite{Gyulassy:1994ew} for HIJING
and in Ref.~\cite{Kisiel:2005hn} for THERMINATOR. For HIJING the events were generated
with default settings and jet quenching on, while for THERMINATOR the default settings were used.
We only focus on the net-charge
distributions within a realistic acceptance of the current experiments at RHIC, that
is pseudorapidity range between -0.5 $<$ $\eta$ $<$ 0.5, transverse momentum range between
0.2 $<$ $p_{T}$ $<$ 2.0 GeV/$c$ with full azimuthal coverage. The analysis is carried out
for 19.6 GeV Au+Au collisions using the events from HIJING model and 200 GeV Au+Au collisions 
using the events from the THERMINATOR model as a function of collision centrality. 
About 5 million events are produced for each centrality studied in both the event generators.
We have checked that the conclusions from each model at other energies are similar to
that presented in this paper. Such a combination of model and beam energy is an arbitrary
choice done to reflect a wide range of kinematics and physics of particle production. The
average charge particle multiplicity counting efficiency is taken to be 65\% following the
efficiency as a function of $p_{T}$ available for charged pions in Ref.~\cite{Abelev:2008ab}.

\section{Bayes method for unfolding of distributions}
The Bayes unfolding algorithm of RooUnfold package is used in general to remove the
effects of measurement resolutions, systematic biases and detection efficiency to
determine the true distributions~\cite{Adye:2011}. The RooUnfoldBayes algorithm based on
Bayes theorem uses the method described by D'Agostini in Ref.~\cite{Agostini:1995yr}.

The procedure of Bayes unfolding can be explained by the causes \textit C and effects
\textit E.  In our study, \textit{causes} correspond to the true multiplicity values and
\textit{effects} to the measured multiplicity values which are affected by the inefficiencies. 
If one observes n(\textit E) events with effect \textit E due to several independent causes $(C_{i}, i = 1, 2, . . . , n_{C} )$
then the expected number of events assignable to each of the causes is given by:

\begin{equation}
\hat n(C_{i}) = n(E) P(C_{i}|E)
\end{equation}
where 
\begin{equation}
P(C_{i}|E) =\frac{P(E|C_{i}) P(C_i)}{\sum_{l=1}^{n_{c}}P(E|C_{l}) P(C_l)} 
\end{equation}

Now if we observe that the outcome of a measurement has several possible effects
$E_{j} (j= 1, 2, 3, ....,n_{E})$ for a given cause $C_{i}$ then the expected number of
events to be assigned to each of the causes and only due to the observed events can be
calculated to each effect by:
\begin{equation}
\hat n(C_{i}) = \sum_{j=1}^{n_{E}}  n(E_{j}) P(C_{i}|E_{j}).
\end{equation}
$P(C_{i}|E_{j})$ is the probability that different causes $C_{i}$ were responsible
for the observed effect $E_{j}$ and is calculated by Bayes theorem as:

\begin{equation}
P(C_{i}|E_{j}) =\frac{P(E_{j}|C_{i}) P_{0}(C_i)}{\sum_{l=1}^{n_{c}}P(E_{j}|C_{l}) P_{0}(C_l)} 
\end{equation}
where $P_{0}(C_{i})$ are the initial probabilities.
If we take into account the inefficiency then the best estimate of the true number of
events is given by,
 \begin{equation}
\hat n(C_{i}) =\frac{1}{\epsilon}_{i}  \sum_{j=1}^{n_{E}}  n(E_{j}) P(C_{i}|E_{j})
{\hspace{.5 in}} {\epsilon}_{i}{\neq}0
\end{equation}
where  ${\epsilon}_{i}$ is the efficiency of detecting the cause $C_{i}$ in any of the
possible effects. If ${\epsilon}_{i} $=0 then  $\hat n(C_{i})$ is set to zero, since the
experiment is not sensitive to the cause $C_{i}$.

The above equation can be written in terms of unfolding or response matrix $M_{ij}$ as,
 \begin{equation}
\hat n(C_{i}) = \sum_{j=1}^{n_{E}} M_{ij}  {n(E_j)}
\end{equation}

The response matrix is constructed by repeated application of Bayes theorem and
the regularization is achieved by stopping the iterations before reaching the "true"
inverse. Further details of the procedure can be found
in~\cite{Agostini:1995yr}.

For the present study, 5M Au+Au collision events are produced for each centrality bin at
$\sqrt{s_{\mathrm {NN}}}$ = 19.6 GeV and 200 GeV using HIJING and
THERMINATOR event generators respectively.
With these events, the $true$ distribution of net-charge $(\Delta N = N ^{+}-N^{-})$ is constructed 
on an event-by-event basis. The positive ($N^+$) and negative ($N^-$) charge particles are selected for each event with
transverse momentum range between 0.2 $<$ $p_{T}$ $<$ 2.0 GeV/$c$ and pseudorapidity
range between -0.5 $<$ $\eta$ $<$ 0.5 with full azimuthal coverage.

The individual \textit{true} $N^+$ and $N^-$ are smeared with a Gaussian function 
with the mean value corresponding to the average efficiency of 65\% as obtained from
the parametrization of the $p_{T}$ dependent efficiency for charged pions from STAR 
experiment~\cite{Abelev:2008ab}. The width of the Gaussian distribution is taken as
10\% of the mean. 
The smeared $N^+$ and $N^-$ distributions will be called
as \textit{measured} distributions. The measured net-charge distribution is then
constructed with these \textit{measured} $N^+$ and $N^-$ distributions.

\bef[h]
\begin{center}
\includegraphics[scale=0.6]{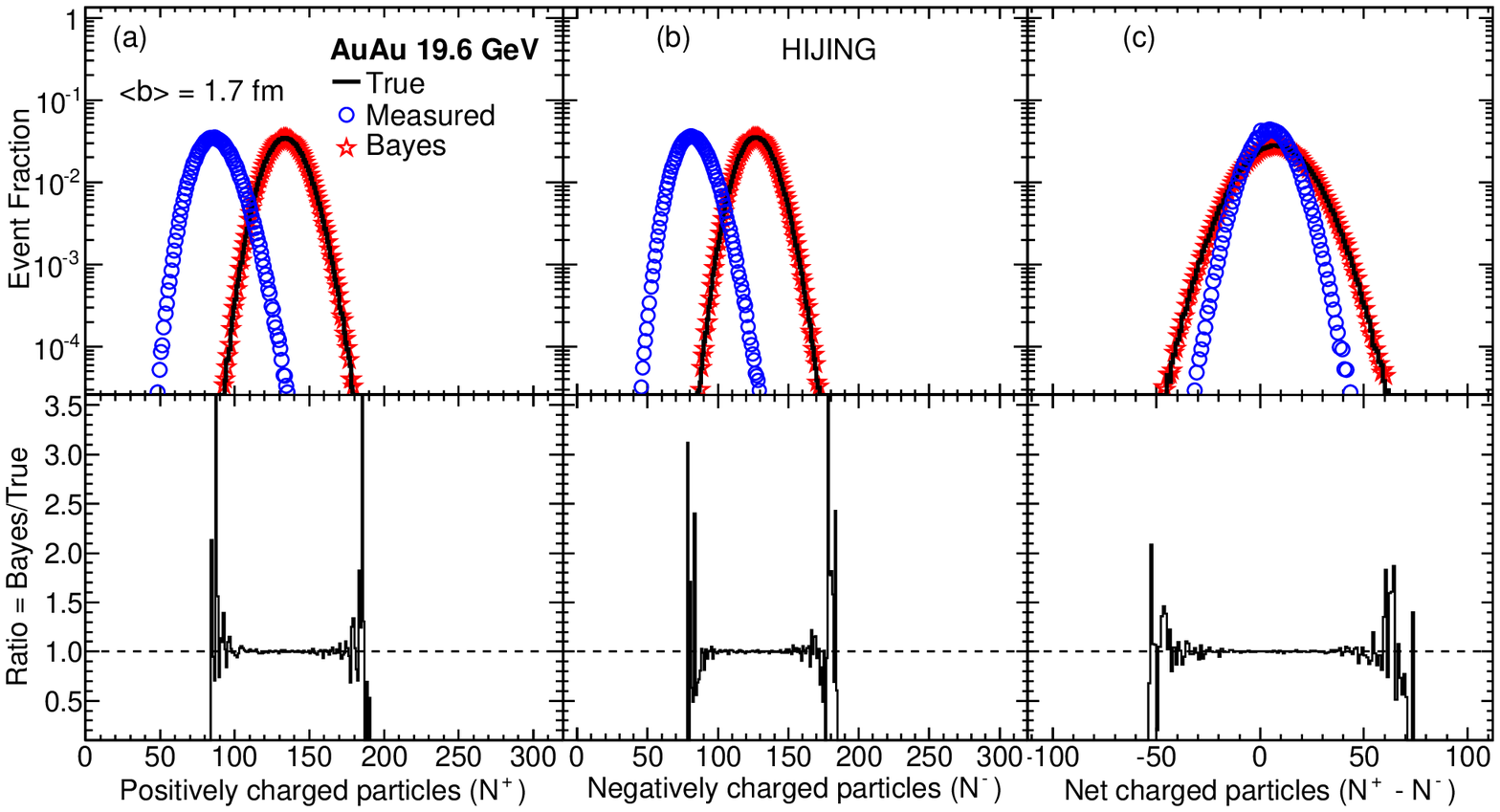}
\caption{(Color online) Top panel: Event-by-event distribution of positive, negative and
net-charge (denoted as ``True'', solid line) in Au+Au collisions for impact parameter $b$
= 1.7 fm at $\sqrt{s_{\mathrm {NN}}}$ = 19.6 GeV from HIJING event generator. Also
shown are the corresponding distributions after applying acceptance and efficiency effects
as discussed in the text (denoted as ``Measured'', open circles). The unfolded
distributions are shown as red stars and denoted as ``Bayes''. Bottom panel: Shows the
ratio of the unfolded to the True distributions.}
\label{fig1}
\end{center}
\eef
To construct the response matrix for each centrality, 2.5M events are used as
\textit{training true} distribution of net-charge and rest of the events are used as
\textit{training measured} (after smearing on an event-by-event basis) distribution. The
events for \textit{training true} and \textit{training measured} are selected separately
to construct the response matrix, in order to avoid the effect of auto-correlation. It
also uses the information of an event that is not measured out of true distributions and
is counted towards the inefficiency while constructing the response matrix.

The measured distribution of net-charge from the remaining 2.5M events is unfolded with 
response matrix obtained from the training procedure using iterative Bayes theorem. 
The number of iterations is called the regularization parameter. The present study uses 
the optimal value of $4$ for the regularization. True, measured and
unfolding are done for finer bins of each centrality and then combined to make 5\% bin to
eliminate the finite centrality bin-width effect. The moments of net-charge distributions
are derived using cumulant method as described in \cite{luo} and are compared for
true, measured and unfolded distributions.

\section{Results and Discussions}

Figure~\ref{fig1} shows the true, measured and unfolded distributions
for positive charge (panel a), negative charge (panel b) and net-charge (panel c) for most central events corresponding
to an average impact parameter of 1.7 fm of Au+Au collisions from HIJING at $\sqrt{s_{\mathrm {NN}}}$ = 19.6 GeV 
on an event-by-event basis. The true distributions are shown as solid lines, measured 
distributions (subjected to particle counting efficiency) are shown as blue open circles 
and the unfolded distributions denoted as ``Bayes''are shown as red star. 
For all the cases, the respective true distributions are reproduced from the measured 
distribution using the unfolding technique. The bottom panel of Fig.\ref{fig1} shows 
the ratio of unfolded to the true distributions corresponding to the same distributions as shown 
in the respective top panels of Fig.\ref{fig1}.
\bef[ht]
\begin{center}
\includegraphics[scale=0.5]{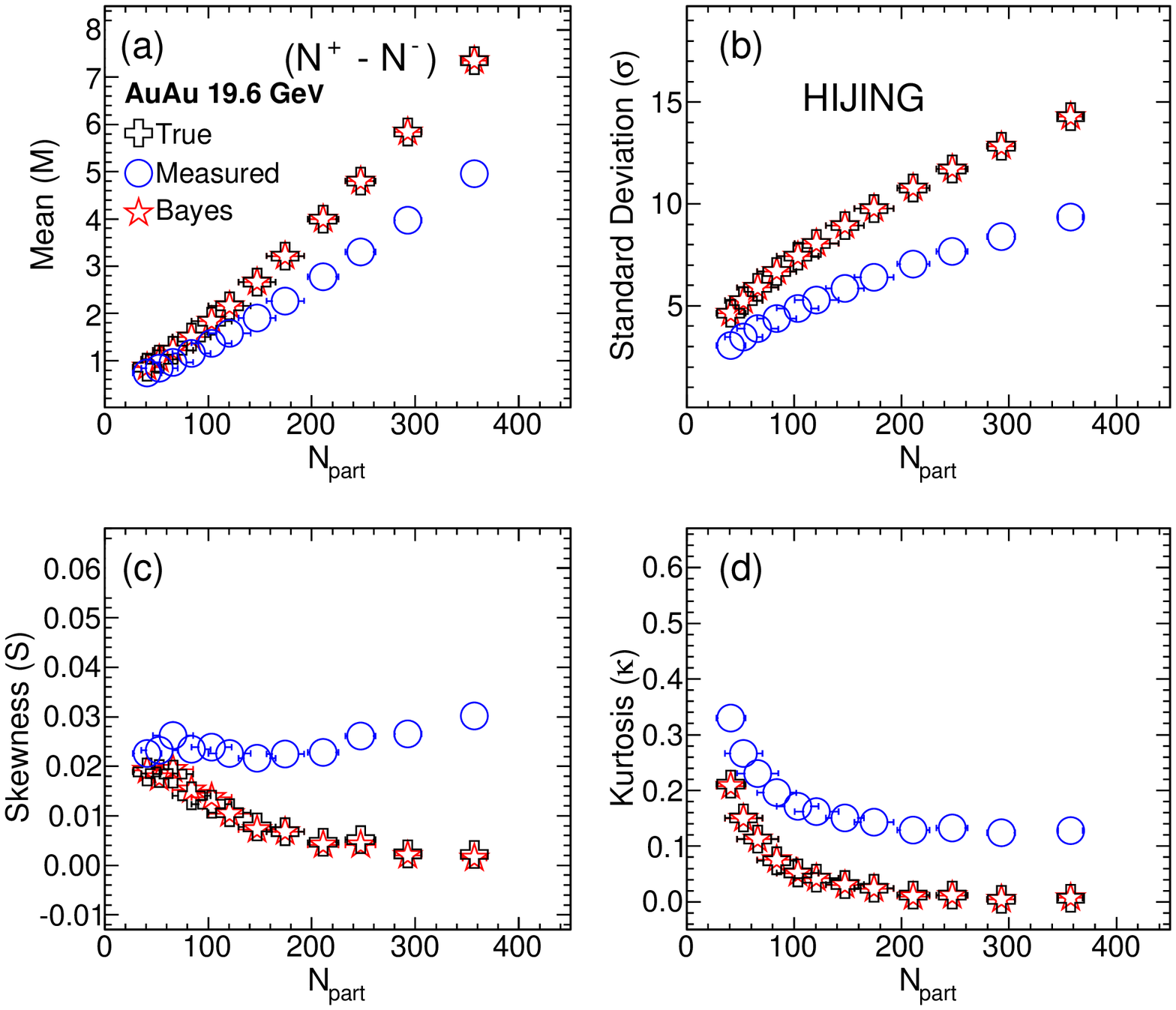}
\caption{(Color online) Mean, standard deviation, skewness and kurtosis of
net-charge distribution in Au+Au collisions at $\sqrt{s_{\mathrm {NN}}}$ = 19.6 GeV from
HIJING event generator. Results are shown for the True, measured  and Bayes unfolded
distributions as a function of $N_{\rm {part}}$.}
\label{fig2}
\end{center}
\eef

The ratio is 
close to unity within the statistical errors, suggesting that the unfolding procedure is able to
get back the true distribution from a measured distribution which is subjected to inefficiencies in particle
counting. Similar conclusions are obtained for distributions for Au+Au collisions at $\sqrt{s_{\mathrm {NN}}}$ = 200 GeV
from THERMINATOR, hence are not shown in this paper. From now onwards we will only concentrate on the net-charge
distributions.

\bef[htp]
\begin{center}
\includegraphics[scale=0.5]{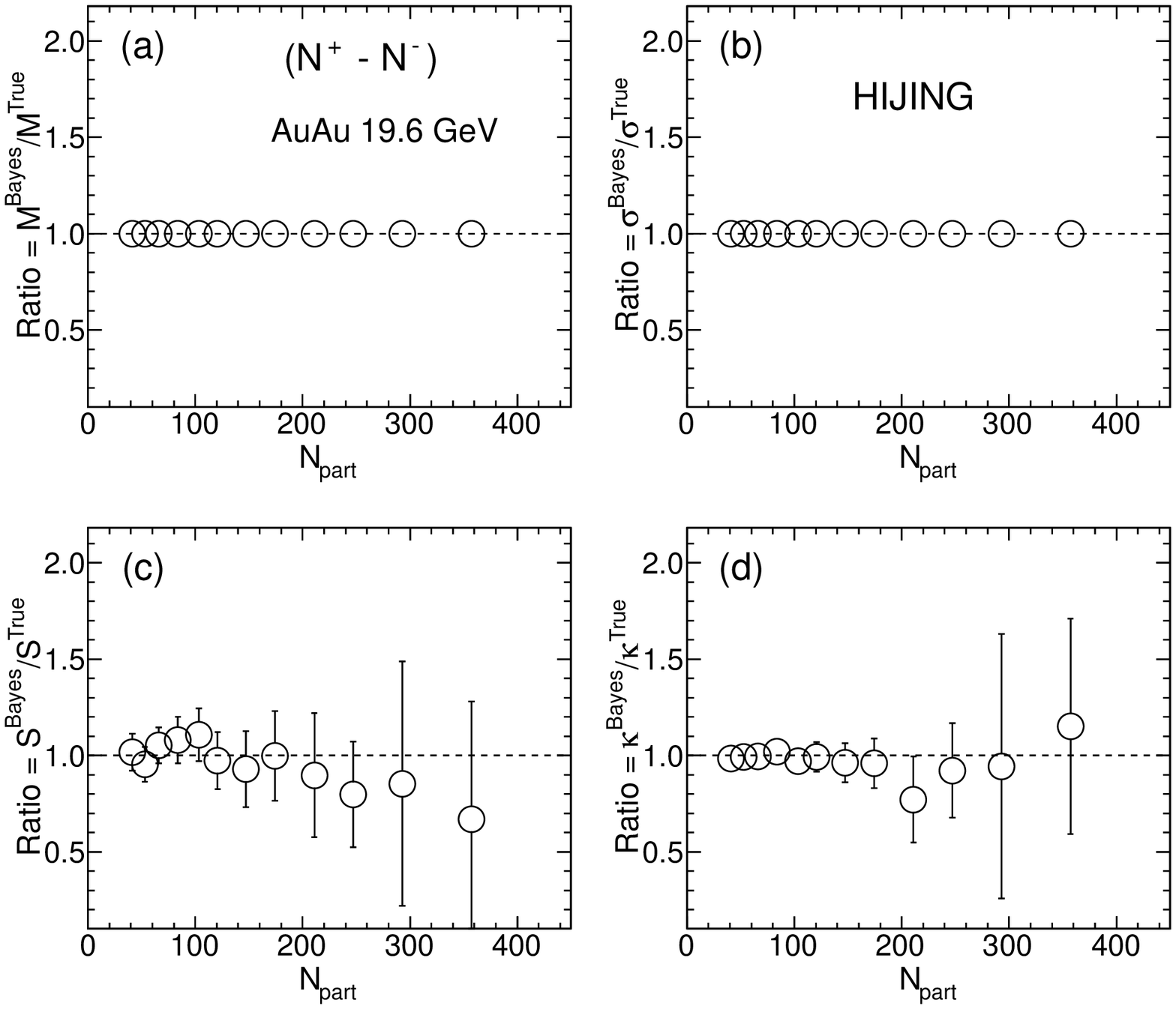}
\caption{Ratio of the unfolded to the True net-charge distribution moments
of Fig.\ref{fig2} as a function of $N_{\rm {part}}$.}
\label{fig3}
\end{center}
\eef
The four moments $M$, $\sigma$, $S$ and $\kappa$ of the net-charge distributions in Au+Au
collisions at $\sqrt{s_{\mathrm {NN}}}$ = 19.6 GeV from the constructed true, measured and
unfolded distributions as a function of centrality ($N_{\rm part}$) are shown in
Fig.\ref{fig2}. The mean and standard deviation increases with $N_{\rm part}$, while
the skewness and kurtosis decreases with $N_{\rm part}$. This is in accordance with the
central limit theorem\cite{Aggarwal:2010wy}. The mean and variance of the measured 
distribution are smaller compared to those of the true, as we have particle counting
inefficiencies for the measured case. The unfolded moments are found to closely follow
the corresponding values of their respective true distributions. 
This can be more clearly seen from the ratio plots in Fig.\ref{fig3}. The value of the
ratio of unfolded to true distribution as a function of $N_{\rm part}$ is around 
unity for all the four moments studied. Thus the unfolding method followed in this paper 
reproduces all the moments of true distribution from the measured distribution. Although
not shown here, similar conclusions are obtained separately for the positive and 
negative charge particle multiplicity distributions.

The centrality dependence of ratio of moments ($\sigma^2/M$) and product of
moments ($S\sigma$ and $\kappa\sigma^2$) are shown in Fig.\ref{fig4}. 
\bef[h]
\begin{center}
\includegraphics[scale=0.7]{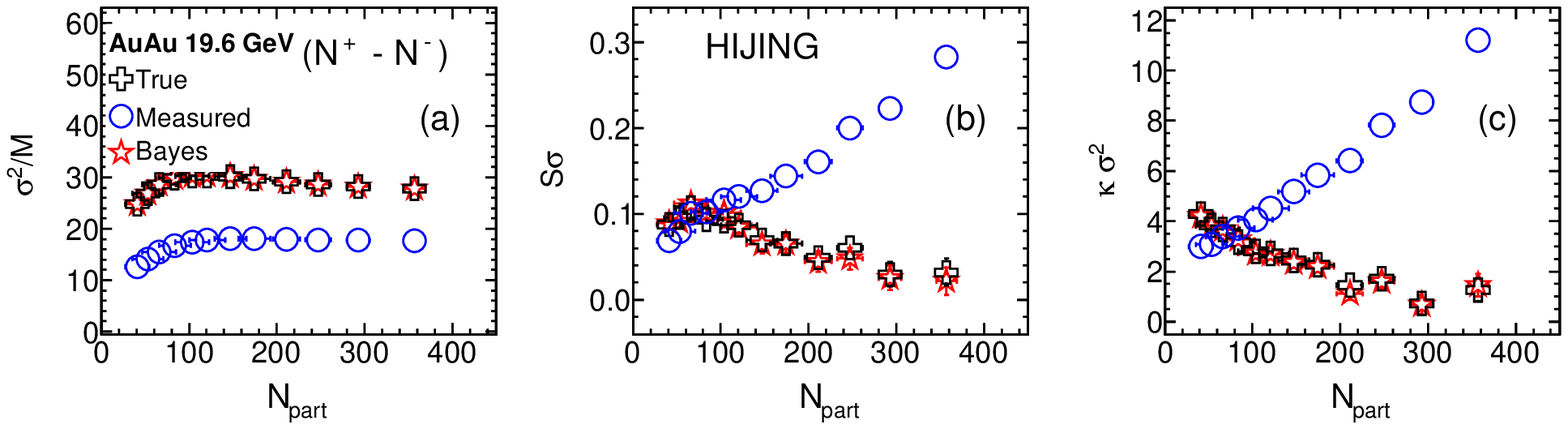}
\caption{(Color online) Ratio (panel a) and product of moments (panel (b) and (c)) of net-charge distributions in Au+Au
collisions at $\sqrt{s_{\mathrm {NN}}}$ = 19.6 GeV from HIJING event generator. The
results are for the True, measured  and Bayes unfolded distributions as a function of $N_{\rm
{part}}$.}
\label{fig4}
\end{center}
\eef
\bef[h]
\begin{center}
\includegraphics[scale=0.5]{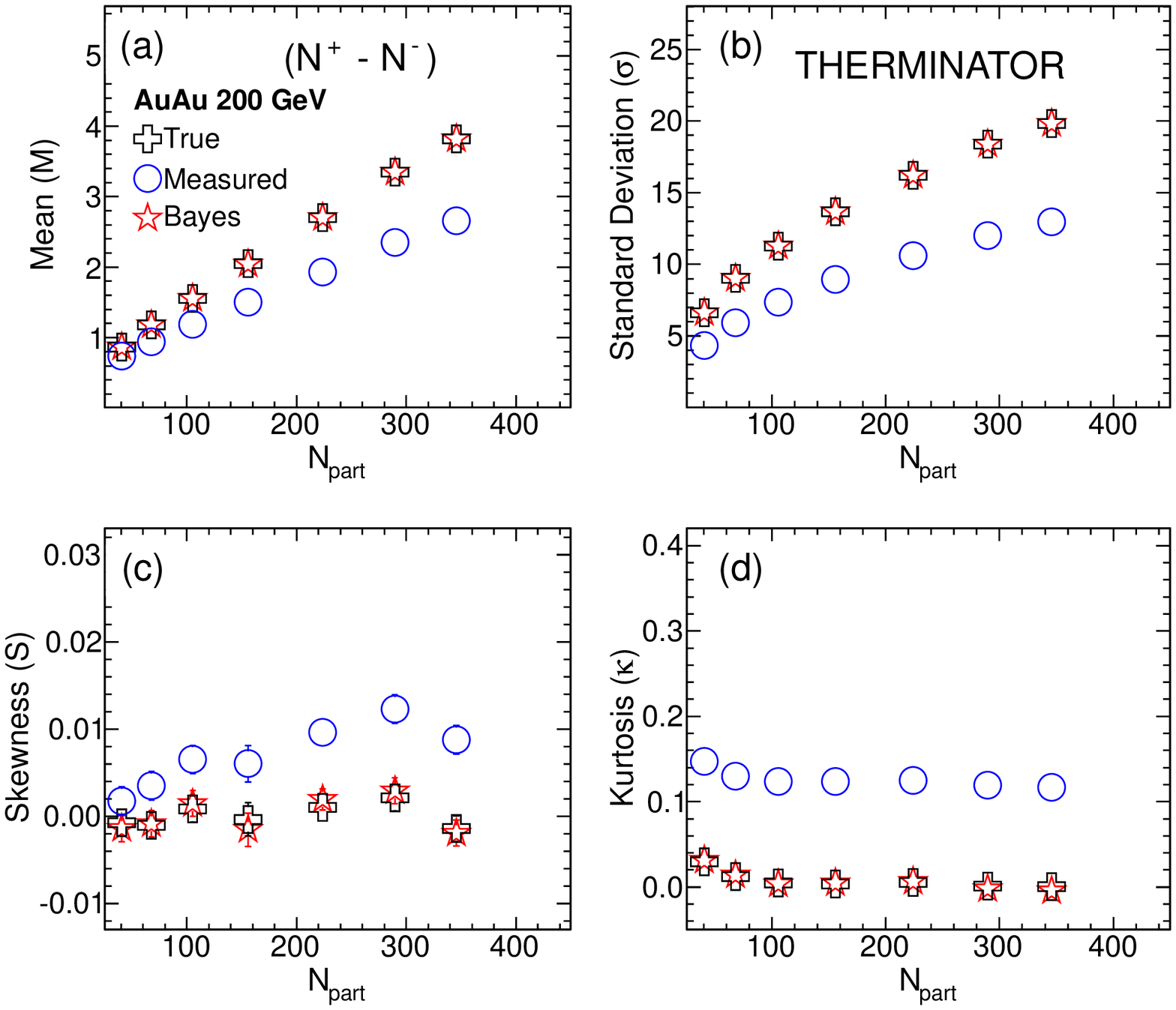}
\caption{(Color online) Mean, standard deviation, skewness and kurtosis of net-charge
distribution in Au+Au collisions at $\sqrt{s_{\mathrm {NN}}}$ = 200 GeV from THERMINATOR
event generator. Results are shown for the True, measured and Bayes unfolded distributions
as
a function of $N_{\rm {part}}$.}
\label{fig5}
\end{center}
\eef
The importance of unfolding is clearly demonstrated by looking at the
dependences of the ratio and product of moments on the $N_{\rm {part}}$. While for
the true distribution the product of moments decreases with $N_{\rm {part}}$,
those for measured actually has an opposite trend. 
Indicating any physics conclusions associated with variation of $S\sigma$ and $\kappa\sigma^2$
with $N_{\rm {part}}$ for net-charge distributions could be highly misleading. 
However, very nice agreement between true and unfolded distributions are observed.
They are nicely consistent even for the product of higher moments ($S\sigma$ and $\kappa\sigma^2$) 
which are very sensitive to the shape of the distributions. Suggesting that
the unfolded distributions are well reproduced as the true distributions by
using Bayes unfolding algorithm.

In order to validate the applicability of unfolding algorithm for different physics
processes, a thermal model based THERMINATOR event generator is also used.
Figure~\ref{fig5} shows the centrality dependence of various moments of net-charge
distribution in Au+Au collisions at $\sqrt{s_{\mathrm {NN}}}$ = 200 GeV from the true,
measured and unfolded distributions. The trends of the moments as a function of 
$N_{\rm {part}}$ is similar to that seen for HIJING (Fig.~\ref{fig2}), although the
magnitude of the moments are different. All the four moments of the unfolded distributions
are well reproduce as the true distributions.

Figure~\ref{fig6} shows the $\sigma^2/M$, $S\sigma$ and $\kappa\sigma^2$ as a function 
of $N_{\rm {part}}$ of net-charge distributions from the true, measured and unfolded 
distributions. Here also, as was seen for the HIJING results (Fig.~\ref{fig4}), the
ratio and products of moments from unfolded distributions are reproduced as true
distributions up to a good extent.
This suggests that the method proposed in this paper works equally well for parent 
distributions produced from very different particle production mechanisms as well
as over a wide range of beam energies. 

\bef[h]
\begin{center}
\includegraphics[scale=0.7]{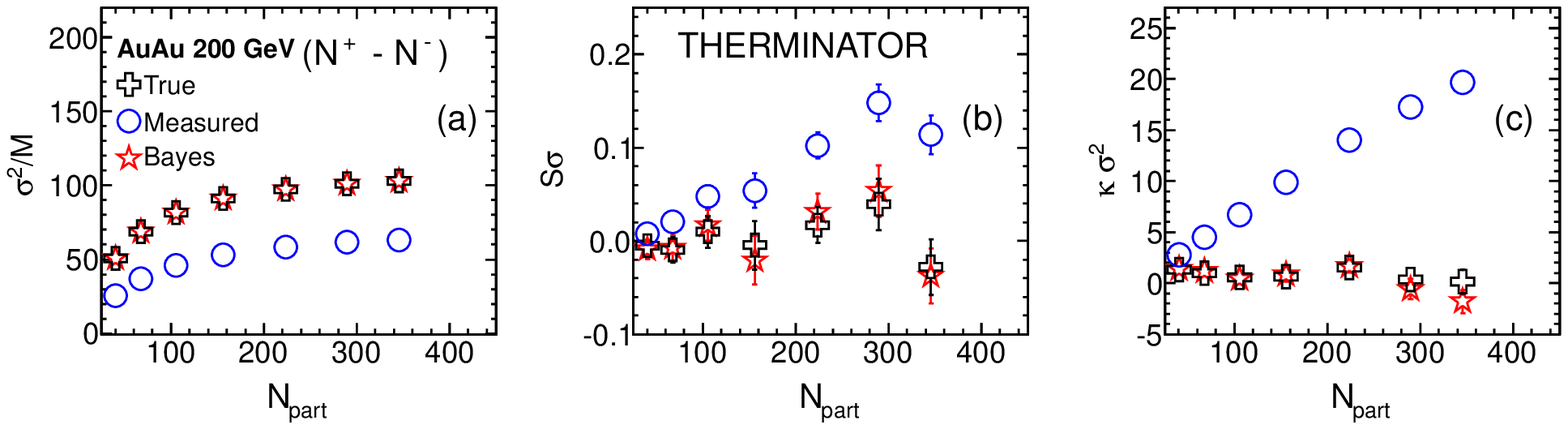}
\caption{(Color online) Product of moments of net-charge distributions in Au+Au collisions
at $\sqrt{s_{\mathrm {NN}}}$ = 200 GeV from THERMINATOR event generator. The results are
for the True, measured and Bayes unfolded distributions as a function of $N_{\rm
{part}}$.}
\label{fig6}
\end{center}
\eef

To study the effect of variation of efficiency on ratio and product of moments, the true
distributions are smeared with a constant efficiency of 65\% to obtain the measured
distributions. Figure~\ref{fig7} and Fig~\ref{fig8} show the $\sigma^2/M$, $S\sigma$ and
$\kappa\sigma^2$ as a function of $N_{\rm {part}}$ of net-charge distributions from the
true, measured and unfolded distributions with constant efficiency in Au+Au collisions at
$\sqrt{s_{\mathrm {NN}}}$ = 19.6 GeV and 200 GeV from HIJING and THERMINATOR event
generators, respectively. Panel (a) of Fig~\ref{fig7} and Fig~\ref{fig8} shows
similar effect as for event-by-event variation of efficiency (panel (a) of Fig~\ref{fig4}
and Fig~\ref{fig6}) on the $\sigma^2/M$ of the measured distributions. The effect of
constant efficiency on $S\sigma$ and $\kappa\sigma^2$ (panel (b) and (c) of Fig~\ref{fig7} and
Fig~\ref{fig8}) of measured distributions is small as compared to event-by-event varying
efficiency.

\bef[ht]
\begin{center}
\includegraphics[scale=0.7]{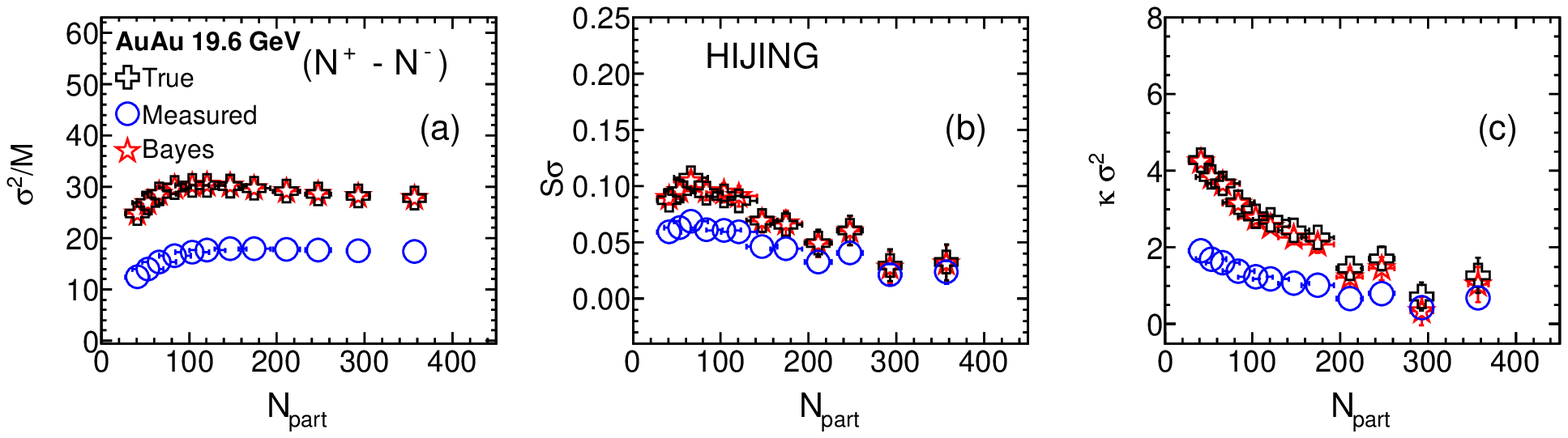}
\caption{(Color online) Product of moments of net-charge distributions in Au+Au collisions
at $\sqrt{s_{\mathrm {NN}}}$ = 19.6 GeV from HIJING event generator with constant
efficiency of 65\%. The results are for the True, measured and Bayes unfolded
distributions as a function of $N_{\rm {part}}$.}
\label{fig7}
\end{center}
\eef

\bef[ht]
\begin{center}
\includegraphics[scale=0.7]{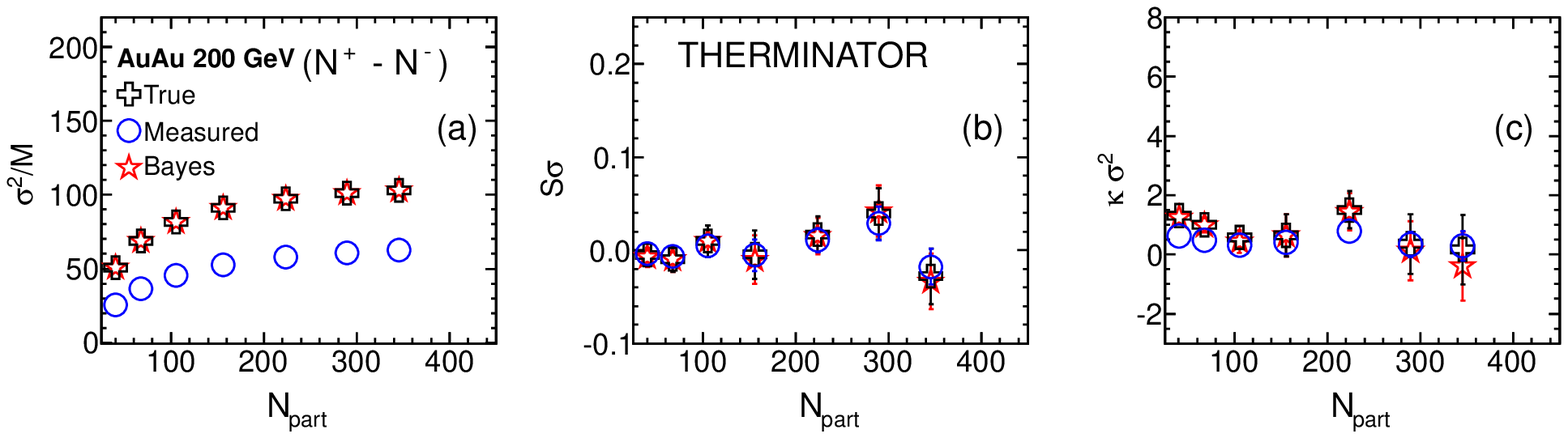}
\caption{(Color online) Product of moments of net-charge distributions in Au+Au collisions
at $\sqrt{s_{\mathrm {NN}}}$ = 200 GeV from THERMINATOR event generator with constant
efficiency of 65\%. The results are for the True, measured and Bayes unfolded
distributions as a function of $N_{\rm {part}}$.}
\label{fig8}
\end{center}
\eef

In order to see the effect of energy dependence on our results we have carried out 
this study for net-charge distributions at midrapidity in 0-5\% central 
Au+Au collisions in HIJING model for $\sqrt{s_{NN}}$ = 19.6, 27, 39, 62.4, 130 and 200 GeV.
The efficiency varies event-by-event as per a Gaussian distribution with mean 
of 65\% and width of 10\% of the mean. 
Figure~\ref{fig9} shows the Mean, standard deviation, skewness and kurtosis for the
above system as a function of beam energy. The mean and variance
of the measured distribution are smaller compared to those of the true, as we have
seen for the centrality dependence study (Fig.~\ref{fig2} and Fig.~\ref{fig5}). This is due 
the particle counting inefficiencies for the measured case. The unfolded moments are found
to closely follow the corresponding values of their respective true distributions.
\bef[ht]
\begin{center}
\includegraphics[scale=0.7]{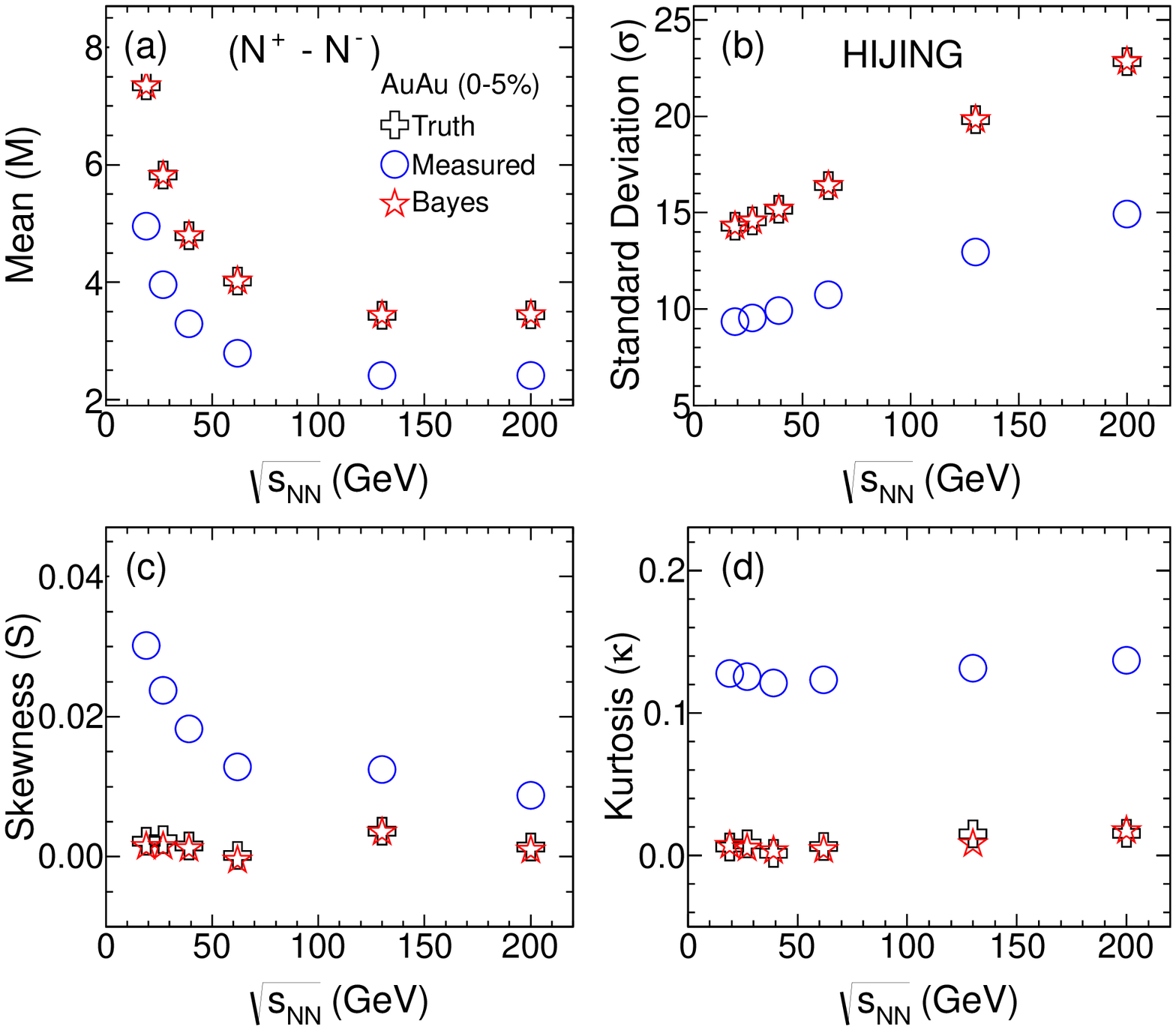}
\caption{(Color online) Mean, standard deviation, skewness and kurtosis of
net-charge distribution in 0-5\% Au+Au collisions as a function of 
 $\sqrt{s_{\mathrm {NN}}}$ from HJING event generator. Results are shown for the 
True, measured  and Bayes unfolded distributions }
\label{fig9}
\end{center}
\eef
Figure~\ref{fig10} shows the $\sigma^2/M$, $S\sigma$ and $\kappa\sigma^2$ as a function 
of $\sqrt{s_{\mathrm {NN}}}$ of 0-5\% Au+Au collisions net-charge distributions 
from the true, measured and unfolded distributions. Here also, as was seen for the 
centrality dependence results, the ratio and products of moments from unfolded distributions 
are reproduced as true distributions up to a good extent.
This suggests that the method proposed in this paper works for parent 
distributions produced over a wide range of beam energies. 

\bef[h]
\begin{center}
\includegraphics[scale=0.7]{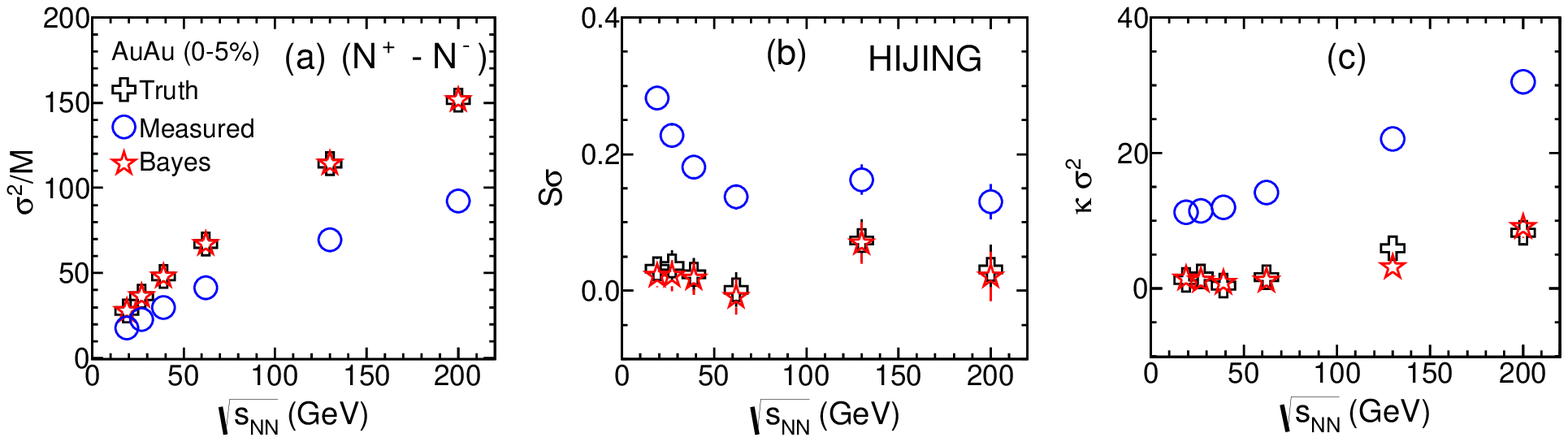}
\caption{(Color online) Product of moments of net-charge distribution in 0-5\% Au+Au collisions
as a function of $\sqrt{s_{\mathrm {NN}}}$ from HIJING event generator. The results are
for the True, measured and Bayes unfolded distributions.}
\label{fig10}
\end{center}
\eef

Our study shows that it is important to correct for event-by-event detector related
effects before proper conclusions can be obtained from higher moments studies
in heavy-ion collisions.We have provided a method of obtaining the true distributions
through an unfolding technique. Such a method keeps the observables same
and hence has the advantage of being used to compare to standard theory calculations. 
Although this procedure can be easily adapted to experimentally measured distributions,
it has two important drawbacks. Unlike the current case, where we have used an event 
generator for the study and the true distribution is available for comparison, in
a real experiment the true distribution is unknown. Hence it is very crucial that a realistic 
modeling of the detector response and particle production is available to obtain the response matrix for the
unfolding calculations. In most cases, the modeling of the particle production and the detector response is highly
event generator dependent and on how realistically the experimental conditions are
simulated. The other disadvantage is, that the procedure works well for high event statistics
as well as high average particle multiplicity per event. We have seen that large uncertainties
enter into the unfolded distributions if we carry out this study with net-protons. 

\section{Summary}

In summary, we have discussed a method to obtain the event-by-event true distributions 
of net-charge from the corresponding measured distributions which are subjected to
detector effects like finite particle counting efficiencies. The approach used is based
on the Bayes method for unfolding of distributions. We have used event generators
HIJING and THERMINATOR to simulate the charged particle distributions produced in 
Au+Au collisions at $\sqrt{s_{NN}}$ = 19.6 to 200 GeV respectively. The charge particle 
counting efficiency was varied by smearing the true distributions, on event-by-event basis
using a Gaussian function with mean 0.65 and width 0.065, to construct the measured
distributions. We have shown that
the unfolded distribution has similar mean, variance, skewness and kurtosis as the
true distributions for all the collision centralities studied. The product of the moments 
 $\sigma^2/M$, $S\sigma$ and $\kappa\sigma^2$ which show an opposite trends versus
$N_{\rm {part}}$ for the measured distributions compared to the true distributions 
are faithfully unfolded back to agree with the true distributions. For cases where
the efficiency of charged particle counting is constant for all events, the differences
between the measured and true are small for the product $S\sigma$ and $\kappa\sigma^2$ 
compared to the ratio  $\sigma^2/M$. The unfolding process is demonstrated to work for 
distributions obtained from widely differing physical mechanism for production of 
charge particle and over a wide range of beam energy and collision centrality. They
also work for the cases where the charged particle counting efficiencies vary 
event-by-event as well as for the case where the efficiencies are constant.

This method has some limitations, in terms of need for a proper 
modeling of the detector response and works well for high multiplicity
and high event statistics dataset. However the main advantage of this
method is that we do not have to construct new observables which cancels out
the detector effects. As the new constructs are usually subjected to difficulties in
physical interpretation and cannot be directly compared to standard theoretical calculations.

\noindent{\bf Acknowledgments}\\
Financial assistance from the Department of Atomic Energy, Government of India is
gratefully acknowledged. BM is supported by the DST SwarnaJayanti project
fellowship. PG acknowledges the financial support from CSIR, New Delhi, India.
\\
%
%

\end{document}